\documentclass[journal]{IEEEtran}

\usepackage{url}

\usepackage{amssymb}
\usepackage{lineno}
\usepackage{tikz}
\usepackage{pgfplots}
\usepackage[utf8]{inputenc}
\usepackage{graphicx}
\usepackage{float}
\usepackage{listings}
\usepackage{verbatim}
\usepackage[shortcuts,acronym]{glossaries}
\usepackage{color}
\usepackage{caption}
\usepackage{booktabs}  
\usepackage{siunitx}
\usepackage{comment}
\usepackage{hyperref}

\usepackage[utf8]{inputenc}

\usepackage[acronym]{glossaries}
\makeglossaries

\usepackage{siunitx} 

\definecolor{mycolor1}{rgb}{0.00000,0.44700,0.74100}
\definecolor{mycolor2}{rgb}{0.85000,0.32500,0.09800}
\definecolor{mycolor3}{rgb}{0.92900,0.69400,0.12500}

\newacronym{CT}{CT}{Computed Tomography}
\newacronym{RNA}{RNA}{Ribonucleic Acid}
\newacronym{COVs}{COVs}{Coronaviruses}
\newacronym{SARS}{SARS}{Severe Acute Respiratory Syndrome}
\newacronym{ARDS}{ARDS}{Acute Respiratory Distress Syndrome}
\newacronym{MERS}{MERS}{Middle East Respiratory Syndrome}
\newacronym{WHO}{WHO}{World Health Organization}
\newacronym{AI}{AI}{Artificial Intelligence}
\newacronym{RT-PCR}{RT-PCR}{ Real-time Reverse Transcription Polymerase Chain Reaction}
\newacronym{RCNN}{RCNN}{Recurrent Convolutional Neural Network}
\newacronym{BCL}{BCL}{Bilateral Convolution Layer}
\newacronym{USM}{USM}{Unsharp Masking}
\newacronym{PC}{PC}{PlantCLEF}
\newacronym{NLM}{NLM}{Non Local Means}
\newacronym{DNLM}{DNLM}{Deceived Non Local Means}
\newacronym{SVM}{SVM}{Support Vector Machine}
\newacronym{LoG}{LoG}{Laplacian of Gaussian}
\newacronym{DoG}{DoG}{Difference of Gaussians}
\newacronym{SP}{SP}{Salt and Pepper}
\newacronym{RPN}{RPN}{Region Proposal Network}
\newacronym{ReLU}{ReLU}{Rectified Linear Unit}
\newacronym{FCN}{FCN}{Fully Convolutional Network}
\newacronym{ROI}{ROI}{Region of Interest}
\newacronym{DDT}{DDT}{Deep Distance Transformer}
\newacronym{Top-WS}{Top-WS}{Watersheds instance segmentation as top-model}
\newacronym{Top-RPN-WS}{Top-RPN-WS}{Region proposal network and watersheds top-model}
\newacronym{Top-Unet3}{Top-Unet3}{U-net top-model}
\newacronym{Top-Unet-ML}{Top-Unet-ML}{U-net with modified loss function top-model}
\newacronym{U-net3}{U-net3}{U-net for three classes}
\newacronym{Unet-ML}{Unet-ML}{U-net with the proposed modified loss function}
\newacronym{DTGT}{DTGT}{Distance Transform Ground-truth}
\newacronym{BTGT}{BTGT}{Border Transform Ground-truth}
\newacronym{WDMC}{WDMC}{Weighted Dice Multiclass Coefficient}
\newacronym{BDE}{BDE}{Boundary Displacement Error}
\newacronym{MAE}{MAE}{Mean Absolute Error}
\newacronym{MSE}{MSE}{Mean Squared Error}
\newacronym{PSP}{PSP}{photostimulable phosphor}
\newacronym{CMOS}{CMOS}{Complementary Metal-oxide-semiconductor}
\newacronym{CMOS-APS}{CMOS-APS}{Complementary Metal Oxide Semiconductor Active Pixel Sensor}
\newacronym{Grad-CAM}{Grad-CAM}{Gradient Class Activation Maps}
\newacronym{PACS}{PACS}{picture archiving and communication system}
\newacronym{TIFF}{TIFF}{tagged image file format}
\newacronym{MRI}{MRI}{Magnetic Resonance Imaging}
\newacronym{CNN}{CNN}{Convolutional Neural Network}
\newacronym{SSDL}{SSDL}{Semi-supervised Deep Learning}
\newacronym{OOD}{OOD}{Out of Distribution}

\newacronym{CAD}{CAD}{Computer Aided Diagnosis}
\newacronym{GAN}{GAN}{Generative Adversarial Network}

\newacronym{PBC}{PBC}{Pseudo-label based Balance Correction}

\begin{document}

\title{ Correcting Data Imbalance  for Semi-Supervised  Covid-19 Detection Using X-ray Chest Images }

\author{\IEEEauthorblockN{Saul~Calderon-Ramirez, Shengxiang-Yang, Armaghan~Moemeni, }
\IEEEauthorblockN{David Elizondo, Simon~Colreavy-Donnelly, Luis~Fernando~Chavarr\'ia-Estrada, Miguel~A.~Molina-Cabello}

\thanks{S.~Calderon-Ramirez, D.~Elizondo, Shengxiang-Yang and S.~Colreavy-Donnelly  work at the Centre for Computational Intelligence (CCI), De Montfort University, United Kingdom  (e-mails:sacalderon@itcr.ac.cr, elizondo@dmu.ac.uk, syang@dmu.ac.uk and simon.colreavy-donnelly@dmu.ac.uk).}
\thanks{S.~Calderon-Ramirez works also at the Instituto Tecnologico de Costa Rica, Costa Rica.}

\thanks{Armaghan~Moemeni works at the School of Computer Science, University of Nottingham, United Kingdom (e-mail:armaghan.moemeni@nottingham.ac.uk).}
\thanks{Luis~Fernando~Chavarria-Estrada works at Im\'agenes M\'edicas Dr Chavarr\'ia Estrada, La Uruca, San Jos\'e, Costa Rica (e-mail:drchavarriaestrada@gmail.com).}

\thanks{Miguel~A. ~Molina-Cabello works at the Department of Computer Languages and Computer Science and the Biomedic Research Institute of M\'alaga (IBIMA),  University of M\'alaga, Spain (e-mail:miguelangel@lcc.uma.es). }

}

\maketitle

\begin{abstract}
The Corona Virus (COVID-19) is an international pandemic that  has quickly propagated throughout the world. A key factor in the fight against this disease is the identification of virus carriers as early and quickly as possible, in a cheap and efficient manner. The application of deep learning for image classification of chest X-ray images of Covid-19 patients, could become a novel pre-diagnostic detection methodology. However, deep learning architectures require large labelled datasets. This is often a limitation when the subject of research is relatively new as in the case of the virus outbreak, where dealing with small labelled datasets is a challenge. Moreover, in the context of a new highly infectious disease, the datasets are also highly imbalanced, with few observations from positive cases of the new disease. In this work we evaluate the performance  of the semi-supervised deep learning architecture known as MixMatch  using a very limited number of labelled observations and  highly imbalanced labelled dataset. Moreover, we demonstrate the critical impact of data imbalance to the model's accuracy. We propose a simple approach for correcting data imbalance, re-weight each observation in the loss function, giving a higher weight to the observations corresponding to the under-represented class. For unlabelled observations, we propose the usage of the pseudo and augmented labels calculated by MixMatch to choose the appropriate weight. The MixMatch method combined with the proposed pseudo-label based balance correction improved classification accuracy by up to 10\%, with respect to the non balanced MixMatch algorithm, with statistical significance. We tested our proposed approach with several available datasets using 10, 15 and 20 labelled observations. Additionally, a new dataset is included among the tested datasets, composed of chest X-ray images of Costa Rican adult patients. 

\end{abstract}

\section{Introduction}

Coronavirus is an endemic kind of virus that affects vertebrate animals, ranging from mammals to reptiles and birds. The SARS-CoV2 virus is a member of this family.  \gls{COVs}  belong to the group of \gls{RNA} viruses. They have the biggest \gls{RNA} genomes found in the viral world, reaching up to 32 KB \cite{science1}. Coronaviruses spread across the gastrointestinal and the respiratory tracks within a large variety of animal groups. The majority of viruses use single animal groups as hosts. However, phylogenetic studies and sequencing of genomes have proven that the \gls{COVs} have managed to migrate  to new host groups \cite{host}, what is referred as a zoonosis.  A zoonosis is a contagious disease produced by an infectious agent, such as a virus, which has managed to move across from a vertebrate animal to humans.  About sixty percent of new infectious diseases are believed to be of zoonosis origin \cite{Jones2008}. Infections caused by zoonosis are of significant concern worldwide. As more and more people regularly travel across the world, the rapid spread is a lurking danger of a worldwide scale.

A key priority for global organizations, including the \gls{WHO} as well as  governments across the world, is to develop tools to enable the identification of virus outbreaks  and to be able to diagnose them in a short time frame. The quick identification of potential virus carriers  is vital to contain a virus outbreak. This is where state of the art \gls{AI} based techniques, such as deep learning, can play a key role, enabling pre-diagnostic and triage systems to effectively identify the presence of the virus in a subject.  They offer quick diagnosis responses to enable health systems to cope with rapid spread of virus out-breaks.

This research extends a novel \gls{SSDL} framework known as  MixMatch \cite{berthelot2019mixmatch} for the detection of COVID-19 based on chest X-ray images. A Semi-supervised learning method allows the combination of labelled and unlabelled data to train the model. This is more cost effective and accessible, as unlabelled data is cheaper than labelled data. Semi-supervised models can easily be adapted for mutations of the virus at a later stage, with relatively small labelled samples. 

We propose a modification for the MixMatch algorithm, designed to improve its accuracy under data imbalance settings.  Added to smaller labelled datasets, in an outbreak situation, datasets can also be strongly imbalanced, as data available for the subjects manifesting symptoms of the new pathogen are more scarce than non-pathogenic patient records. 

\subsection{Use of X-ray images towards the diagnosis of COVID-19}

A common, well established and robust method for the detection of COVID-19 virus is the \gls{RT-PCR} test \cite{chan2020improved}. This is a molecular test, which uses respiratory tract samples to identify and confirm infection of COVID-19 \cite{whopcr}. The objective of the method is to find the nucleic acid of the SARS-CoV-2 within both the lower and the upper respiratory areas. Samples from symptomatic patients suspected of infection of the COVID-19 are gathered \cite{vashist2020vitro}. However, new research shows the need for testing asymptomatic individuals as well \cite{bai2020presumed}. \gls{RT-PCR} is the main method used for detecting the presence of the disease \cite{ai2020correlation}.  Nevertheless, the costs  associated to the use of \gls{RT-PCR} can be significant, since the facilities and trained personnel needed to perform these tests can be expensive. These severely limit the use of this technique in less industrialized countries, making urgent the need to develop more accessible methods, adding the possible need of testing asymptomatic patients. \cite{narayanan2020pooling}. 

Diagnosing COVID-19 based on medical imaging can be a reliable and accurate alternative, and is still under exploration. The accuracy and sensitivity levels of this approach as a first stage in COVID-19 detection using chest images, have been analyzed in a number of studies \cite{chung2020ct,fang2020sensitivity}. 

 The usage of X-ray images for COVID-19 diagnosis has been studied recently. In \cite{borghesi2020} the authors proposed a severity score using radiography chest images. The dataset used in this study had a total of 783 SARS-CoV-2 infected cases. The score was used to identify patients that could potentially acquire more life threatening symptoms. Several studies  \cite{chen2020epidemiological,chung2020ct,song2020emerging} have suggested that in a small number of people there is a low level of sensitivity towards the manual detection of alterations using medical images of the chest which can  indicate the presence of COVID-19. The use of features extracted and learned by a machine might overcome the variable subjective evaluation of X-ray images.   This leads us to explore the potential implementation of deep learning solutions using more widely available and less expensive chest X-ray images. As typical deep learning architectures require many labelled images, we aim to explore the usage of \gls{SSDL} for COVID-19 detection using X-ray images, evaluating it under another frequent challenge; data imbalance.

\subsection{Contribution}

In this paper, we extensively test the \gls{SSDL} technique known as MixMatch \cite{berthelot2019mixmatch} in a variety of data imbalance situations, with a very limited number of labelled observations. We aim to assess MixMatch's performance under real-world scenarios, specifically medical imaging in the context of a virus out-break, where small labelled samples are available with a strong under-representation of the new pathology, leading to imbalanced datasets. An imbalanced dataset can frequently lead also to a distribution mismatch between the labelled and unlabelled dataset, as described in \cite{oliver2018realistic}. 

Moreover, in this work we propose a simple, yet effective approach for correcting data imbalance for the \gls{SSDL} algorithm MixMatch. We implement a loss based imbalance correction, giving more weight to the under-represented classes in the labelled dataset, a common approach for this aim. In the context of MixMatch, we  make use of  the pseudo-label and augmented labels predictions to choose the corresponding class-weight. The implemented \gls{SSDL} solution for COVID-19 detection makes use of unlabelled data. This might help improve model's accuracy, in the absence of high quality labelled data.

The proposed method uses chest X-ray images. X-ray machines are commonly available, which results in a wealth of unlabelled datasets due to the shortage of radiologists and technicians who can label the images. As an example, India, with its current 1.44 billion population, has a ratio between radiologists and patients of 1:100,000 \cite{Arora2014}.  However, X-ray machines can be found even in remote areas in under-developed countries, compared to other medical devices like computer tomography scanners \cite{shah2015assessment}. 

We also make available a first sample of a chest-X ray dataset from the Costa Rican medical private clinic Imagenes Medicas Dr. Chavarria Estrada,  with observations containing no findings, and test its usage for training the  \gls{SSDL} framework. 

In the event of a viral outbreak, it becomes essential to help health practitioners to quickly identify and classify viral pathologies using digital X-ray images. Outbreaks create a large number of cases, which require the intervention of trained radiologists. Labeling data is time consuming, and in the context of a virus out-break gathering high quality and reliable labelled data can be challenging. \gls{SSDL} can provide much needed key support for the diagnosis, trace and isolation of the COVID-19 infection and other future pandemics through an early, fast and cheap diagnosis, by using  more widely available unlabelled data.

\section{Related Work}

\subsection{Deep learning for Chest X-ray based  COVID-19 detection}

 The identification of COVID-19 infection based on X-ray images is a new challenge. Thus, up to date there is not much research available with regards to the use of deep learning models for automatically identifying COVID-19 infection. This is the reason why this paper presents mainly  pre-published work in the area up-to-date. Since most pre-published articles have not been peer reviewed, it is used here as a general guide and not as a reference towards performance. 
 
A classification model based on a support vector machine fed with deep features was presented in \cite{sethy2020detection}. Different common deep learning architectures were used for feature extraction. These included: VGG16, AlexNet, GoogleNet , VGG19, several variations of Inception and Resnet, DenseNet201 and XceptionNet. The dataset used included a total of fifty observations with half representing COVID-19 images and the other half representing a combination of pneumonia and normal images. The COVID-19 images were acquired from the Github repository created by Dr. Joseph Cohen from the University of Montreal \cite{cohen2020covid}. COVID-19 negative images were downloaded from the public repository on X-ray images presented in  \cite{kermany2018identifying}. The highest level of accuracy was obtained with the ResNet50 model which was combined with a support vector machine as a top model. An accuracy of around 95\%, with statistical significance, was obtained.

Several machine learning algorithms were compared in \cite{covid-xray1}. Some of the methods considered included: support vector machines, random forests and \gls{CNN} models. The results reported the \gls{CNN} model as the best performing approach, with an accuracy of 95.2\%. The dataset used in this work includes 48 Cases for COVID-19$^{+}$ and 23 for negative COVID-19 cases from  Dr. Cohen's  repository \cite{cohen2020covid}. Data augmentation was used to deal with scarce labelled data.

Another study involving the use of \gls{CNN}s along with transfer-learning for the automatic classification of pneumonia, COVID-19 and images presenting no lung pathology was presented in \cite{Apostolopoulos2020}. The authors used a 10-fold cross-validation, to test the following \gls{CNN} architectures: VGG-19, MobileNet v2, Inception, Xception and Inception ResNet v2. An accuracy of around 93\% was obtained in the identification of COVID-19, with the use of a VGG-19 model. No statistical significance tests were performed.  As for the data used in \cite{Apostolopoulos2020},  similar to related proposed solutions,  positive COVID-19 cases were extracted from \cite{cohen2020covid}, while pneumonia and no lung pathology observations were taken from   \cite{kermany2018identifying}. 

A deep learning model for the automatic detection of COVID-19 and pneumonia was proposed in \cite{Chowdhury2020}. The system proposed classifies images into three classes;  COVID-19$^{+}$, viral  pneumonia and normal readings.  To increase the number of observations,  the authors relied on data augmentation techniques including rotation, translation and scaling, along with transfer-learning. The architectures tested included: AlexNet, ResNet19, DenseNet201 and SqueezeNet. A combination of the datasets from \cite{cohen2020covid} was used in this research. The SqueezeNet model outperforms all the other \gls{CNN} networks. Regarding the data used in such work, a combination of two data repositories \cite{wang2017chestx,kermany2018identifying} was used for viral and normal image categories, and the data repository in \cite{cohen2020covid} was used for positive COVID-19 cases. 

Explainability for deep learning models  is an important feature for medical imaging based systems \cite{holzinger2019causability}. Model uncertainty estimation is a common approach to enforce model explainability and usage safety \cite{holzinger2019causability}. A COVID-19 detection system with uncertainty assessment was proposed in \cite{Ghoshal2020}. By providing practitioners with a confidence factor of the prediction, the overall reliability of the system is improved. A high correlation between the prediction accuracy of the model and the level of uncertainty was reported  \cite{Ghoshal2020}. The dataset used for positive COVID-19 cases also uses Dr. Cohen's repository \cite{cohen2020covid}, and normal X-ray readings were collected from \cite{kermany2018identifying}.

In \cite{khobahi2020coronet}, a semi-supervised approach for defining relevant features for COVID-19 detection was developed. The suspicious regions were extracted by training a semi-supervised auto-encoder architecture that minimizes the reconstruction error. This approach relies in the wider availability of COVID-19$^{-}$ cases to learn relevant features.   Such extracted features were used for classifying the input observations into three classes; COVID-19$^{+}$, pneumonia and normal, using a common supervised \gls{CNN} approach. The extracted features were used to enforce model explainability. Similar to previous reviewed approaches, the datasets provided in \cite{cohen2020covid,kermany2018identifying}  were used. 

Similarly, the work in \cite{cohen2020predicting} used a feature extractor built from training a model to classify X-ray images in larger datasets with non COVID-19 observations. The model was trained for the regression of COVID-19 severity.   Similar to \cite{khobahi2020coronet}, the feature extractors built ease the extraction of further information from the model,  improving the model's explainability. A wider range of datasets were used in such work for training the feature extractor \cite{demner2016preparing,bustos2019padchest,johnson2019mimic,majkowska2020chest,wang2017chestx,irvin2019chexpert}. 

In summary, the reviewed works implemented transfer-learning and data augmentation to deal with limited labelled data. Fewer works trained more specific feature extractors \cite{cohen2020predicting,khobahi2020coronet}.  The datasets in \cite{cohen2020covid,wang2017chestx,kermany2018identifying}  have been used extensively in  previous work. The frequently used dataset in  \cite{cohen2020covid} 
includes COVID-19$^{+}$ observations  made available by Dr. Joseph Cohen, from the University of Montreal \cite{cohen2020covid}. The images were collected from journal websites such as \url{radiopaedia.org}, the Italian Society of Medical and Interventional Radiology. The images were also collected from recent publications in this area such as  \cite{cohen2020covid}. The dataset is made of chest X-ray images involving over 100 patients. Their ages range from 27 to 85 years old. The countries of origin include: Iran, China, Italy, Taiwan, Australia, Spain and the United Kingdom. A warning has been raised by the authors on \cite{cohen2020covid} with regards to any diagnostic performance claims prior to doing a proper clinical study. As for the dataset available in \cite{kermany2018identifying},  frequently used in previous work for normal and pneumonia readings, all of them correspond  to samples taken from pediatric Chinese patients.  The usage of such data as negative COVID-19 cases can be less reliable, since   different populations were sampled for COVID-19 and no COVID-19 cases.  Observations of adults (with ages ranging between 20 and 86 years old) were used for COVID-19$^{+}$ cases, while for the normal and pneumonia cases in \cite{kermany2018identifying}, the images were sampled from pediatric patients.
Therefore, in this work we test a wider variety of sources for COVID-19$^{-}$ cases, including a new dataset with Costa Rican adult patients.

Little exploration on the benefits of using  a fully  \gls{SSDL} model can be found in the literature. Furthermore, to our knowledge no work on the impact and correction of data imbalance in \gls{SSDL} for COVID-19 detection has been developed so far in the literature.  

\subsection{Semi-supervised deep learning and data imbalance correction}

In general deep learning models require a large number of labelled observations to provide good levels of generalisation. This limitation makes it hard to implement these techniques to medical applications since there is a lack of labelled data \gls{SSDL} is gaining increasing popularity in the academic community. It is well suited to deal with datasets which are poorly labelled, or have few labels,  making \gls{SSDL} attractive for computer aided medical imaging analysis.

Semi-supervised methods require the use of both labelled $S_{l}=(X_{l},Y_{l})$ and unlabelled samples $S_u = X_{u} = \left\{ \boldsymbol{x}_{1},\ldots,\boldsymbol{x}_{n_{u}}\right\}$. Each labelled observation in $X_{l}=\left\{ \boldsymbol{x}_{1},\ldots,\boldsymbol{x}_{n_{l}}\right\}$ has an associated label in the set $Y_{l}=\left\{ y_{1},\ldots,y_{n_{l}}\right\} $. No labels are associated to the unlabelled set.

\gls{SSDL} architectures can be classified as follows: Pre-training, self-training (also known as pseudo-labelled) and regularization based. Some of the regularization methods include generative based approaches, along consistency loss term as well as graph based. An extensive survey on \gls{SSDL} approaches can be found in \cite{van2020survey}.

The MixMatch approach developed in \cite{berthelot2019mixmatch} merged intensive data augmentation with unsupervised regularization and pseudo-labelled based semi-supervised learning. This method produced better results compared to other regularized,  pseudo-labelled and generative based \gls{SSDL} methods as shown in  \cite{berthelot2019mixmatch}. 

Data imbalance in the labelled dataset, can be approached as a particularisation of the data distribution mismatch problem outlined in \cite{oliver2018realistic}, when the unlabelled dataset presents a different distribution. This is common under real-world usage conditions of \gls{SSDL} techniques. In  \cite{oliver2018realistic},  authors made a first glance at the impact of \gls{OOD} data in the unlabelled dataset $S_u$, leading to a distribution mismatch between the distributions of $S_l$ and $S_u$. 

The work in \cite{calderonramirez2020mixmood} went deeper into the impact of \gls{OOD} data in \gls{SSDL}. Authors tested several distribution mismatch scenarios with different \gls{OOD} data contamination degrees, and different \gls{OOD} data sources. The results showed an important influence on the degree of \gls{OOD} data in the unlabelled dataset $S_u$, as also the distribution of the \gls{OOD} observations by itself.

In \cite{hyun2020class}, authors explored further the impact of the distribution mismatch, in the particular case of using imbalanced datasets. The results showed a classification error rate decrease, ranging from 2\% to 10\%  for the \gls{SSDL} model. Furthermore, the authors proposed a straightforward approach for correcting such accuracy degradation. The approach assigned weights to each unlabelled observation, depending on the number of observations per class. Higher weights were used for under-represented observations in the unlabelled loss term $\mathcal{L}_u$.  To pick the right weight for each unlabelled observation, the highest label predicted with the model yielded for the current epoch, was used. The authors implemented and tested the approach in the mean teacher model \cite{tarvainen2017mean}. The results demonstrated a significant accuracy gain by implementing the proposed approach. We base our contribution on these findings, and propose an extended data imbalance correction approach into MixMatch in the context of semi-supervised COVID-19 detection.

\subsection{MixMatch}\label{subsec:MixMatch}

The proposed \gls{SSDL} method is based on the MixMatch \cite{berthelot2019mixmatch} algorithm. It creates a set of pseudo-labels, and also implements an unsupervised regularization term. The consistency loss term used by the MixMatch method minimizes the distance between the pseudo-labels and  predictions that the model makes on the unlabelled dataset $X_{u}$.

The average model output of a transformed input $x_{j}$ was used to estimate pseudo-labels:
\begin{equation} 
\widehat{\boldsymbol{y}}{}_{j}=\frac{1}{K}\sum_{\eta=1}^{K}f_{\overrightarrow{w}}\left(\Psi^{\eta}\left(\boldsymbol{x}_{j}\right)\right)
\end{equation}

Here $K$ corresponds to the number of transformations (like image flipping) $\Psi^{\eta}$  performed. Based on the work by \cite{berthelot2019mixmatch}, a value of $K=2$ is recommended. The authors also mentioned that the  estimated pseudo-label
$\widehat{\boldsymbol{y}}{}_{j}$ usually presents a high entropy value. This can increase the number of  non-confident estimations. Therefore, the output array $\widehat{\boldsymbol{y}}$ was sharpened with a temperature $\rho$:
\begin{equation} 
    s\left(\widehat{\boldsymbol{y}},\rho\right)_{i}=\frac{\widehat{y}_{i}^{1/\rho}}{\sum_{j}\widehat{y}_{j}^{1/\rho}}
\end{equation}

When  $\rho\rightarrow0$, the sharpened distribution $\widetilde{\boldsymbol{y}}=s\left(\widehat{\boldsymbol{y}},\rho\right)$ becomes a Dirac function, assuming a one-hot vector representation. The term $\widetilde{S}_{u}=\left(X_{u},\widetilde{Y}\right)$ defines the dataset with the sharpened estimated pseudo labels. It is assumed here that $\widetilde{Y}=\left\{ \widetilde{\boldsymbol{y}}_{1},\widetilde{\boldsymbol{y}}_{2},\ldots,\widetilde{\boldsymbol{y}}_{n_{u}}\right\}$

In \cite{berthelot2019mixmatch} the authors argued that data augmentation is a key aspect when it comes to \gls{SSDL}. The authors used the MixUp approach, as proposed in \cite{zhang2017mixup}, to further augment  data using both labelled and unlabelled observations:
\begin{equation}
\left(S'_{l},\widetilde{S}'_{u}\right)=\Psi_{\textrm{MixUp}}\left(S_{l},\widetilde{S}_{u},\alpha\right)
\end{equation}

The MixUp method proposed to create new observations based on a linear interpolation of a combination of unlabelled (together with their pseudo-labels) and labelled data. More specifically, for two labelled or pseudo labelled data pairs $\left(\boldsymbol{x}_{a},y_{a}\right)$ and $\left(\boldsymbol{x}_{b},y_{b}\right)$, MixUp creates a new observation with its corresponding label $\left(\boldsymbol{x}',y'\right)$ based on the following steps:

\begin{enumerate}

\item  Sample the MixUp parameter $\lambda$ based on a Beta distribution $\lambda\sim\textrm{Beta}\left(\alpha,\alpha\right)$.

\item  Make sure that $\lambda>0.5$. This is done by making $\lambda'=\max\left(\lambda,1-\lambda\right)$

\item  Produce a new observation based on a lineal interpolation of the two observations: $\boldsymbol{x}'=\lambda'\boldsymbol{x}_{a}+\left(1-\lambda'\right)\boldsymbol{x}_{b}$.

\item  Generate the corresponding pseudo-label for the new observation $y'=\lambda'y_{a}+\left(1-\lambda'\right)y_{b}$.
\end{enumerate}

The augmented datasets $ \left(S'_{l},\widetilde{S}'_{u}\right)$ were used by the MixMatch algorithm to train  a model as specified in the training function $T_{\textrm{MixMatch}}$:
\begin{equation}
f_{\overrightarrow{w}}=T_{\textrm{MixMatch}}\left(S_{l},X_{u},\alpha,\lambda\right)=\underset{\boldsymbol{w}}{\textrm{argmin}}\mathcal{L}\left(S,\boldsymbol{w}\right)
\end{equation}
\[
\mathcal{L}\left(S,\boldsymbol{w}\right)=\sum_{\left(\boldsymbol{x}_{i},\boldsymbol{y}_{i}\right)\in S'_{l}}\mathcal{L}_{l}\left(\boldsymbol{w},\boldsymbol{x}_{i},\boldsymbol{y}_{i}\right)+
\]
\begin{equation}
\gamma r(t) \sum_{\left(\boldsymbol{x}_{j},\widetilde{\boldsymbol{y}}_{j}\right)\in\widetilde{S}'_{u}}\mathcal{L}_{u}\left(\boldsymbol{w},\boldsymbol{x}_{j},\widetilde{\boldsymbol{y}}_{j}\right)
\end{equation}

For the labelled loss term, a cross-entropy loss was used; $\mathcal{L}_{l}\left(\boldsymbol{w},\boldsymbol{x}_{i},\boldsymbol{y}_{i}\right)=\delta_{\textrm{cross-entropy}}\left(\boldsymbol{y}_{i},f_{\boldsymbol{w}}\left(\boldsymbol{x}_{i}\right)\right)$. As for the unlabelled loss term, an  Euclidean distance was implemented  $\mathcal{L}_{u}\left(\boldsymbol{w},\boldsymbol{x}_{j},\widetilde{\boldsymbol{y}}_{j}\right)=\left\Vert \widetilde{\boldsymbol{y}}_{j}-f_{\boldsymbol{w}}\left(\boldsymbol{x}_{j}\right)\right\Vert$.  The coefficient $r(t)$ was proposed  as a ramp-up function that increases its value as the epochs $t$ increase. In our implementation,  $r(t)$ was set to $t/3000$. The $\gamma$ factor was used as a regularization weight. This coefficient  controls the influence on unlabelled data. It is important to highlight that unlabelled data has also an effect on the \textit{labelled} data term $\mathcal{L}_{l}$. The reason being that unlabelled data is used   to artificially increase data observations by using the MixUp method for also the labelled term.

\section{Proposed Method: Pseudo-label based balance correction}
In this work an implementation of a data imbalance correction in the loss function of the MixMatch method is proposed. Positive results were yielded in \cite{hyun2020class} for correcting dataset imbalance by weighting the unsupervised loss function terms in a per observation basis. The authors in \cite{hyun2020class} developed a similar approach by modifying the \gls{SSDL} framework known as  mean teacher  \cite{tarvainen2017mean}.   We extend this approach for the MixMatch algorithm, but using both the  pseudo-labels and augmented labels for selecting the  appropriate weights for both the unlabelled and labelled loss terms. We refer to the proposed approach in this work as \gls{PBC}.

The number of observations per class is used to compute the array of correction coefficients $\textbf{c}$. The actual computation is done by calculating the  array $\textbf{v}$ using the inverse of the amount of observations available in each class $S_l$: $v_i =\frac{1}{n_i}$. Here $n_i$ corresponds to the total amount of observations for class $i$. The next step consists in  the computation of the array with the normalized weights $\textbf{c}$ as follows: 
\begin{equation}
c_{i}=\frac{v_{i}}{\sum_{j}^{C}v_{j}}
\end{equation}

Where $C$ corresponds to the total number of classes, where in this work $C=2$, as a binary classification model is developed.  The augmented, pseudo, and original labels $\textbf{y}_i$ and $\widetilde{\textbf{y}}_{j}$, are contained in the augmented labelled and unlabelled datasets, $S'_l$ and $\widetilde{S}'_{u}$, respectively, after the MixUp method mentioned in Section \ref{subsec:MixMatch} is executed. Such augmented labels  are  used to select its corresponding weight in $\textbf{c}$. To do so,   the one-hot vector notation of the labels is converted  to a numeric one:
\begin{equation}
b_{i}=\underset{k}{\textrm{argmax}}y_{k,i}
\end{equation}
\begin{equation}
\widetilde{b}_{j}=\underset{k}{\textrm{argmax}}\widetilde{y}_{k,j}
\end{equation}

for every $b_i$ and $\widetilde{b}_{j}$ observation in $S'_l$ and $\widetilde{S}'_{u}$, respectively.

Both the loss function and the calculated weights are used to weight both loss terms:
\[
\mathcal{L}\left(S,\boldsymbol{w}\right)=\sum_{\left(\boldsymbol{x}_{i},\boldsymbol{y}_{i}\right)\in S'_{l}}\mathcal{L}_{l}\left(\boldsymbol{w},\boldsymbol{x}_{i},\boldsymbol{y}_{i},b_j\right)+
\]
\begin{equation}
\gamma r(t) \sum_{\left(\boldsymbol{x}_{j},\widetilde{\boldsymbol{y}}_{j}\right)\in\widetilde{S}'_{u}}\mathcal{L}_{u}\left(\boldsymbol{w},\boldsymbol{x}_{j},\widetilde{\boldsymbol{y}}_{j},\widetilde{b}_{j}\right)
\end{equation}

The chosen indices are used in the array of weights $\textbf{c}$. We used a cross-entropy and mean squared error loss for the labelled and unlabelled loss terms, respectively. Therefore, the modified cross-entropy and MSE functions are respectively described as follows: $\mathcal{L}_{l}\left(\boldsymbol{w},\boldsymbol{x}_{i},\boldsymbol{y}_{i}\right)=\delta_{\textrm{cross-entropy}}\left(c_{b_{i}}\boldsymbol{y}_{i},c_{b_{i}}f_{\boldsymbol{w}}\left(\boldsymbol{x}_{i}\right)\right)$ and $\mathcal{L}_{u}\left(\boldsymbol{w},\boldsymbol{x}_{j},\widetilde{\boldsymbol{y}}_{j}\right)=\left\Vert c_{\widetilde{b}_{j}}\widetilde{\boldsymbol{y}}_{j}-c_{\widetilde{b}_{j}}f_{\boldsymbol{w}}\left(\boldsymbol{x}_{j}\right)\right\Vert$. The numerical estimated and real labels are then used for indexing the array $\textbf{c}$. The  re-weighted loss functions are minimized as usual \footnote{All code, experimental scripts and results  is temporally  available at \url{shorturl.at/stI49}. Upon paper publication, we are going to make it available through a public github repository.}.

\section{Datasets}
A system to classify x-ray images into: COVID-19+ and no lung pathology (COVID-19-) is presented in this work. We used different previously existing datasets, and add the usage of a new one, containing negative COVID-19 cases. 

The following previously existing datasets were used in this work: 

\begin{enumerate}
    \item  \textbf{COVID-19$^{+}$ dataset:} Images containing COVID-19$^{+}$ observations were collected from the publicly available github repository accessible from \cite{cohen2020covid}. This repository was built by Dr. Joseph Cohen, from the University of Montreal \cite{cohen2020covid}. The images were collected from journal websites such as \url{radiopaedia.org}, the Italian Society of Medical and Interventional Radiology. Images were also collected from recent publications in this area such as  \cite{cohen2020covid}. Only images containing signs of COVID-19$^{+}$ were used in this study. All other images relating to \gls{MERS}, \gls{ARDS} and \gls{SARS} were discarded.  This reduced the dataset to a subset containing 102 front chest X-ray containing COVID-19$^{+}$ observations. The gray-scaled observations were stored  with varying resolutions from $400 \times 400$ up to $2500 \times 2500$ pixels. 
    
    \item \textbf{Chinese pediatric patients dataset:} A dataset of 5856 observations containing images of pneumonia and normal observations was defined in  \cite{kermany2018identifying}. The patient sample used for the study correspond to Chinese children \cite{kermany2018identifying}. These images are divided into 4273 observations of pneumonia (including viral and bacterial) and 1583 of observations with no lung pathology (normal).  We used the observations with no findings, and refer to it as the Chinese pediatric dataset. The negative and pneumonia observations from this dataset have been used extensively in recent related research to COVID-19 detection \cite{Narin,Zhang,Wang,El-DinHemdan,Salman2020,Apostolopoulos2020}.  Most of the images were stored with a resolution of $1300 \times 600$ pixels. 
    
    \item \textbf{ChestX-ray8 dataset:} The ChestX-ray8 dataset, made available in \cite{irvin2019chexpert}, is also used for the category of no findings in this work. The dataset includes  224,316 chest radiographs from 65,240 patients from Stanford Hospital, US. The studies were done between October 2002 and July 2017. We picked a sample of this dataset available in its website\footnote{\url{https://www.kaggle.com/nih-chest-xrays/sample/data}} given the low labelled data setting used in this work. Patients sampled in this dataset were aged from 0 to 94 years old.

    \item \textbf{Indiana Chest X-ray  dataset:} The dataset published in \cite{demner2016preparing}   gathers 8121  images from the Indiana Network for Patient Care. Only the observations with no pathologies were used in this work.  The dataset can be accessed from its repository\footnote{\url{https://www.kaggle.com/raddar/chest-xrays-indiana-university}}. Images were stored with a  resolution of $1400 \times 1400$ pixels.

\end{enumerate}

In this work we also used a dataset we gathered from a Costa Rican private clinic, Clinica Imagenes Medicas Dr. Chavarria Estrada. The data corresponds to chest X-rays from 153 different patients, with ages ranging from 7 to 86 years old. 63\% of the patients were female and 37\% are male. The images were taken using a Konica Minolta digital X-ray machine with 0.175 of pixel spacing. The images were stored with a resolution of $1907 \times 1791$ pixels. As the images were digitally sampled, no tags or manual labels are contained in the images\footnote{The dataset  is temporally available at \url{shorturl.at/dghsQ}. We will move it to a public repository upon paper publication.}. 

All the datasets have been preprocessed to exclude  artifacts (manual labels), in the cases where one of them does not present any, to avoid artifact bias. Data augmentation using flips and rotations is implemented. No crops were used to avoid losing  regions that might be important for image discrimination. Images stored with 8 bits were replicated by 3 to use the selected \gls{CNN} architecture.

\section{Experiments definition}

We used the following hyper-parameters used for the MixMatch model for all the experiments performed: $K=2$ transformations, $T=0.5$ of sharpening temperature and $\alpha=0.75$ for the beta distribution\footnote{ The  MixMatch implementation used in this work is based on the implementation available in repository \url{https://github.com/noachr/MixMatch-fastai}}.    A Wide-ResNet  \cite{zagoruyko2016wide} model has been used for all the experiments, with an input image size of $110\times110$ pixels, and the following hyper-parameters: a weight decay of 0.0001, a learning rate of 0.00001, a batch size of 12 observations, a cross-entropy loss function and an adam optimizer with a 1-cycle policy \cite{smith2018disciplined}. 

For each configuration, we trained the model 10 times for a total of 50 epochs. For each run, a sample dataset of 204 observations was picked from both the evaluated COVID$19^{-}$ dataset  and the COVID-$19^{+}$ dataset available in \cite{cohen2020covid}. Therefore, a total of 10 different training and test samples were used. The same samples were used for all the tested algorithm variations.  A completely balanced validation dataset comprising the 30\% of the 204 observations was used. 

To assess the data imbalance impact, we evaluated both the supervised and the semi-supervised architectures using three balance configurations: 50\%50\%, 80\%/20\% and 70\%/30\% for the labelled dataset $S_l$. The under-represented class corresponds to the COVID-19$^{+}$ class. We tested different sizes of labelled samples, $n_l=10$, $n_l=15$ and $n_l=20$. The remaining data was used as unlabelled data, with close to a 50\% data balance between the two classes.  This leads to a distribution mismatch between  $S_u$ and $S_l$. Tables \ref{table:costarica}, \ref{table:china}, \ref{table:NIS} and \ref{table:indiana} show this layout. Given the low labelled setting, we report the highest validation accuracy, assuming the usage of  early stopping to avoid over-fitting.  We trained the MixMatch model with both the uncorrected loss function and the proposed \gls{PBC} modification for data imbalance correction. For reference, we also  tested the supervised model with balance correction and without it. 

Table \ref{table:summary} summarizes the accuracy gains when using MixMatch with \gls{PBC} vs. not using MixMatch, and using MixMatch with no balance correction (under the same balance conditions) vs. using MixMatch with \gls{PBC}. A non-parametric Wilcoxon test was performed to detect whether the accuracy gain was statistically significant (with $p>0.1$) across the 10 runs (observations) sampled. Gains not statistically significant according such criteria are written in italic in Table \ref{table:summary}.

Finally, as a qualitative experiment, we calculated the gradient activation maps using the technique proposed in \cite{selvaraju2017grad}\footnote{We used the FastAI implementation available of the gradient activation maps available in \url{https://forums.fast.ai/t/gradcam-and-guided-backprop-intergration-in-fastai-library/33462}}. For this qualitative experiment we compared the supervised model and the MixMatch modification with the proposed \gls{PBC}. The objective of this experiment was to spot the changes on the regions used by the model to output its decision, when trained with the semi-supervised approach. A sample with 20 labelled observations and around 180 unlabelled observations (for the MixMatch model with \gls{PBC})  was used for training the model. A completely balanced dataset of 61 observations was used for validation.  We trained a Densenet121 model for 50 epochs, for both the supervised and semi-supervised frameworks. Figure \ref{fig:HeatMapsINDIANA} includes sampled heatmaps for the chest X-ray8 and Indiana datasets. The net weights in the final output layer for each entry, and the real and predicted labels are also shown for each output image in Figure \ref{fig:HeatMapsINDIANA}.

\begin{table*}
\centering
\begin{tabular}{c|c|c|c|cc|cc|cc}
\hline
\textbf{SSDL} & \textbf{COVID-19-} & \textbf{COVID-19+} & \textbf{LB} & \multicolumn{2}{c|}{\textbf{$n_l =10$}} & \multicolumn{2}{c|}{\textbf{$n_l=15$}} & \multicolumn{2}{c}{\textbf{$n_l=20$}} \\
              &                    &                    &             & $\overline{x}$                & $s$                & $\overline{x}$                & $s$               & $\overline{x}$                & $s$               \\ \hline
No            & 50\%               & 50\%               & NA          & 0.871              & 0.039              & 0.912              & 0.049             & 0.951              & 0.025             \\ \cline{2-10} 
              & 70\%               & 30\%               & Yes         & 0.877              & 0.04               & 0.9                & 0.053             & 0.931              & 0.034             \\
              &                    &                    & No          & 0.877              & 0.04               & 0.924              & 0.056             & 0.931              & 0.044             \\ \cline{2-10} 
              & 80\%               & 20\%               & Yes         & 0.876              & 0.06               & 0.903              & 0.058             & 0.922              & 0.037             \\
              &                    &                    & No          & 0.876              & 0.079              & 0.907              & 0.072             & 0.938              & 0.035             \\ \hline
Yes           & 50\%               & 50\%               & NA          & 0.941              & 0.035              & 0.955              & 0.025             & 0.957              & 0.03              \\ \cline{2-10} 
              & 70\%               & 30\%               & Yes         & 0.955              & 0.027              & 0.947              & 0.035             & 0.95               & 0.029             \\
              &                    &                    & No          & 0.907              & 0.042              & 0.9                & 0.049             & 0.914              & 0.028             \\ \cline{2-10} 
              & 80\%               & 20\%               & Yes         & 0.957              & 0.025              & 0.964              & 0.021             & 0.96               & 0.02              \\
              &                    &                    & No          & 0.922              & 0.031              & 0.926              & 0.047             & 0.919              & 0.033             \\ \hline
\end{tabular}
\caption{Accuracy results with the COVID-$19^{-}$ from the Costa Rican dataset, the higher, the better. LB stands for label balancing, with usual weight correction for the supervised model, and the proposed \gls{PBC} for the MixMatch model. A total of 10, 15 and 20 labelled observations were tested. Two data imbalance settings were tested, with 70\%/30\% and 80\%/20\%. The sample mean $\overline{x}$ and the sample standard deviation $s$ are reported.  }
\label{table:costarica}
\end{table*}

\begin{table*}[]
\begin{centering}
\begin{tabular}{c|c|c|c|cc|cc|cc}
\hline
\textbf{SSDL} & \textbf{COVID-19-} & \textbf{COVID-19+} & \textbf{LB} & \multicolumn{2}{c|}{\textbf{$n_l =10$}} & \multicolumn{2}{c|}{\textbf{$n_l=15$}} & \multicolumn{2}{c}{\textbf{$n_l=20$}} \\
              &                    &                    &             & $\overline{x}$                & $s$                & $\overline{x}$                & $s$               & $\overline{x}$                & $s$               \\ \hline
No            & 50\%               & 50\%               & NA          & 0.882              & 0.077              & 0.868              & 0.08              & 0.925              & 0.039             \\ \cline{2-10} 
              & 70\%               & 30\%               & Yes         & 0.812              & 0.5                & 0.815              & 0.089             & 0.883              & 0.048             \\
              &                    &                    & No          & 0.823              & 0.048              & 0.815              & 0.087             & 0.868              & 0.064             \\ \cline{2-10} 
              & 80\%               & 20\%               & Yes         & 0.857              & 0.107              & 0.898              & 0.052             & 0.93               & 0.053             \\
              &                    &                    & No          & 0.823              & 0.125              & 0.872              & 0.066             & 0.93               & 0.037             \\ \hline
Yes           & 50\%               & 50\%               & NA          & 0.945              & 0.036              & 0.95               & 0.026             & 0.963              & 0.028             \\ \cline{2-10} 
              & 70\%               & 30\%               & Yes         & 0.925              & 0.042              & 0.93               & 0.053             & 0.943              & 0.034             \\
              &                    &                    & No          & 0.902              & 0.058              & 0.898              & 0.091             & 0.915              & 0.044             \\ \cline{2-10} 
              & 80\%               & 20\%               & Yes         & 0.947              & 0.037              & 0.957              & 0.022             & 0.962              & 0.028             \\
              &                    &                    & No          & 0.847              & 0.122              & 0.857              & 0.141             & 0.895              & 0.042             \\ \hline
\end{tabular}
\par\end{centering}
\caption{Accuracy results with the COVID-$19^{-}$ cases gathered from the Chinese pediatric repository available in \cite{kermany2018identifying}. LB stands for label balancing, with usual weight correction for the supervised model, and the proposed \gls{PBC} for the MixMatch model. A total of 10, 15 and 20 labelled observations were tested.  Two data imbalance settings were tested, with 70\%/30\% and 80\%/20\%. The sample mean $\overline{x}$ and the sample standard deviation $s$ are reported.}
\label{table:china}
\end{table*}

\begin{table*}[]
\begin{centering}
\begin{tabular}{c|c|c|c|cc|cc|cc}
\hline
\textbf{SSDL} & \textbf{COVID-19-} & \textbf{COVID-19+} & \textbf{LB} & \multicolumn{2}{c}{\textbf{$n_l =10$}} & \multicolumn{2}{c|}{\textbf{$n_l=15$}} & \multicolumn{2}{c}{\textbf{$n_l=20$}} \\
              &                    &                    &             & $\overline{x}$                & $s$                & $\overline{x}$                & $s$               & $\overline{x}$                & $s$               \\ \hline
No            & 50\%               & 50\%               & NA          & 0.756              & 0.062              & 0.727              & 0.062             & 0.756              & 0.05              \\ \cline{2-10} 
              & 70\%               & 30\%               & Yes         & 0.732              & 0.039              & 0.723              & 0.043             & 0.752              & 0.038             \\
              &                    &                    & No          & 0.739              & 0.051              & 0.744              & 0.053             & 0.773              & 0.049             \\ \cline{2-10} 
              & 80\%               & 20\%               & Yes         & 0.729              & 0.051              & 0.721              & 0.054             & 0.768              & 0.047             \\
              &                    &                    & No          & 0.735              & 0.052              & 0.739              & 0.07              & 0.777              & 0.05              \\ \hline
Yes           & 50\%               & 50\%               & NA          & 0.803              & 0.059              & 0.814              & 0.052             & 0.84               & 0.038             \\ \cline{2-10} 
              & 70\%               & 30\%               & Yes         & 0.816              & 0.048              & 0.815              & 0.038             & 0.839              & 0.079             \\
              &                    &                    & No          & 0.782              & 0.054              & 0.76               & 0.068             & 0.782              & 0.051             \\ \cline{2-10} 
              & 80\%               & 20\%               & Yes         & 0.798              & 0.05               & 0.818              & 0.044             & 0.824              & 0.039             \\
              &                    &                    & No          & 0.735              & 0.056              & 0.74               & 0.075             & 0.752              & 0.048             \\ \hline
\end{tabular}
\par\end{centering}
\caption{Accuracy results with the COVID-$19^{-}$ cases gathered from the ChestX-ray8  repository available in \cite{irvin2019chexpert}. LB stands for label balancing, with usual weight correction for the supervised model, and the proposed \gls{PBC} for the MixMatch model. A total of 10, 15 and 20 labelled observations were tested.   The sample mean $\overline{x}$ and the sample standard deviation $s$ were  reported.}
\label{table:NIS}
\end{table*}

\begin{table*}[]
\begin{centering}
\begin{tabular}{c|c|c|c|cc|cc|cc}
\hline
\textbf{SSDL} & \textbf{COVID-19-} & \textbf{COVID-19+} & \textbf{LB} & \multicolumn{2}{c|}{\textbf{$n_l =10$}} & \multicolumn{2}{c|}{\textbf{$n_l=15$}} & \multicolumn{2}{c}{\textbf{$n_l=20$}} \\
              &                    &                    &             & $\overline{x}$                & $s$                & $\overline{x}$                & $s$               & $\overline{x}$                & $s$               \\ \hline
No            & 50\%               & 50\%               & NA          & 0.845              & 0.044              & 0.853              & 0.053             & 0.879              & 0.038             \\ \cline{2-10} 
              & 70\%               & 30\%               & Yes         & 0.834              & 0.042              & 0.839              & 0.053             & 0.874              & 0.046             \\
              &                    &                    & No          & 0.845              & 0.058              & 0.86               & 0.05              & 0.869              & 0.061             \\ \cline{2-10} 
              & 80\%               & 20\%               & Yes         & 0.845              & 0.048              & 0.829              & 0.053             & 0.856              & 0.042             \\
              &                    &                    & No          & 0.84               & 0.041              & 0.827              & 0.045             & 0.853              & 0.066             \\ \hline
Yes           & 50\%               & 50\%               & NA          & 0.905              & 0.047              & 0.918              & 0.038             & 0.908              & 0.029             \\ \cline{2-10} 
              & 70\%               & 30\%               & Yes         & 0.882              & 0.067              & 0.902              & 0.046             & 0.902              & 0.042             \\
              &                    &                    & No          & 0.837              & 0.078              & 0.819              & 0.109             & 0.834              & 0.037             \\ \cline{2-10} 
              & 80\%               & 20\%               & Yes         & 0.86               & 0.076              & 0.889              & 0.056             & 0.885              & 0.035             \\
              &                    &                    & No          & 0.803              & 0.062              & 0.747              & 0.095             & 0.795              & 0.078             \\ \hline
\end{tabular}
\par\end{centering}
\caption{Accuracy results with the COVID-$19^{-}$ cases gathered from  Indiana
dataset \cite{demner2016preparing}. LB stands for label balancing, with usual weight correction for the supervised model, and the proposed \gls{PBC} for the MixMatch model. A total of 10, 15 and 20 labelled observations were tested.  The sample mean $\overline{x}$ and the sample standard deviation $s$ were  reported  }
\label{table:indiana}
\end{table*}

\begin{table*}[]
\begin{centering}
\begin{tabular}{c|c|c|c|c|c|c}
\hline
\textbf{Dataset}  & \textbf{COVID-19-} & \textbf{COVID-19+} & \textbf{Comparison} & \textbf{$n_l =10$} & \textbf{$n_l=15$} & \textbf{$n_l=20$} \\ \hline
Costa Rican       & 70\%               & 30\%               & MM+PBC vs. No MM    & +0.07              & +0.046            & +0.018            \\
                  &                    &                    & MM+PBC vs. MM       & +0.048             & +0.046            & +0.036            \\
                  & 80\%               & 20\%               & MM+PBC vs. No MM    & +0.081             & +0.06             & +0.038            \\
                  &                    &                    & MM+PBC vs. No MM    & +0.034             & +0.038            & +0.041            \\ \hline
Chinese pediatric & 70\%               & 30\%               & MM+PBC vs. No MM    & +0.113             & +0.115            & +0.06             \\
                  &                    &                    & MM+PBC vs. MM       & \textit{+0.023}    & +0.031            & \textit{+0.028}   \\
                  & 80\%               & 20\%               & MM+PBC vs. No MM    & +0.09              & +0.058            & +0.031            \\
                  &                    &                    & MM+PBC vs. No MM    & +0.1               & +0.099            & +0.066            \\ \hline
Chest-Xray8       & 70\%               & 30\%               & MM+PBC vs. No MM    & +0.083             & +0.092            & +0.087            \\
                  &                    &                    & MM+PBC vs. MM       & \textit{+0.033}    & +0.055            & +0.057            \\
                  & 80\%               & 20\%               & MM+PBC vs. No MM    & +0.069             & +0.096            & +0.056            \\
                  &                    &                    & MM+PBC vs. No MM    & +0.063             & +0.0774           & +0.072            \\ \hline
Indiana           & 70\%               & 30\%               & MM+PBC vs. No MM    & +0.048             & +0.063            & +0.027            \\
                  &                    &                    & MM+PBC vs. MM       & +0.045             & +0.082            & +0.067            \\
                  & 80\%               & 20\%               & MM+PBC vs. No MM    & +0.014             & +0.059            & \textit{+0.029}   \\
                  &                    &                    & MM+PBC vs. No MM    & +0.056             & +0.141            & +0.09             \\ \hline
\end{tabular}
\par\end{centering}
\caption{Accuracy gain comparison when using no \gls{SSDL} (No MM) vs. MixMatch with the proposed loss balancing correction (MM+\gls{PBC}), and to using Mix Match with no balancing correction (MM) vs. MixMatch with the proposed loss balancing correction (MM+\gls{PBC}). The accuracy gain is evaluated for the tested number of labelled observations (10, 15 and 20). Italic entries correspond to non statistically meaningful gains, after performing a Wilcoxon test, with $p > 0.1$. }
\label{table:summary}
\end{table*}


\section{Results and analysis}

The results using  accuracy as a metric for the Costa Rican dataset are depicted in Table \ref{table:costarica}. The base-line accuracy is rather high for very limited labelled settings, even with the base-line supervised model, with accuracies ranging from 87\% to 95\%, using 10 and 20 labels, respectively. \gls{SSDL} is perhaps only attractive when using 10 labels, with an accuracy gain of around 7\%, as displayed in the summary Table \ref{table:summary}.  The accuracy gain from implementing \gls{PBC} vs. using the non-balanced MixMatch approach remains similar in disregard of the number of labels used, always with statistical significance. However, the accuracy gain of using MixMatch, even with the \gls{PBC} modification, diminishes as the number of labels increases. The accuracy gain is rather similar for both of the data imbalance configurations tested. As seen in Table \ref{table:costarica}, the implemented  \gls{PBC} corrects the data imbalance impact, yielding similar results when using the completely balanced dataset. 
 
Regarding the test results using the Chinese pediatric dataset, the base-line supervised  accuracy results are initially low (from 86\% to 92\%), giving more room for \gls{SSDL} accuracy gain, as seen in Table \ref{table:china}. The usage of MixMatch with the proposed \gls{PBC} over regular supervised learning yields an accuracy gain over +11\% as seen in Table \ref{table:summary}. Similar to the Costa Rican dataset, as the number of labels increases, the accuracy gain decreases. The benefit of using the \gls{PBC} over the off-the-shelf MixMatch implementation is higher when facing a more imbalanced dataset scenario, as seen in Table \ref{table:summary} for the Chinese dataset. The accuracy gain is almost three times higher when using the 80\%/20\% configuration, increasing from around +3\% to +10\%, for the 70\%/30\% and 80\%/20\% imbalance scenarios, respectively. The \gls{PBC} is able to almost correct the impact of data imbalance, as its accuracy shown in Table \ref{table:china} often is similar to the base-line MixMatch accuracy with a  balanced dataset. 

Table \ref{table:NIS} summarizes the results yielded for the Chest X-ray8  dataset. The base-line accuracy for the supervised model is the lowest from the tested datasets, sitting at  around 75\%. The accuracy gain of using MixMatch with \gls{PBC} versus the usual supervised model ranges from +5\%  to +9.6\%, as seen in Table \ref{table:summary}, in the row for the Chest X-ray8 dataset. As for the accuracy gain of using MixMatch with \gls{PBC} vs. MixMatch with no balance correction, it stays around +3 to +5\%  for the 70\%/30\%  imbalance configuration. Higher accuracy gains are obtained when dealing with the more challenging imbalance scenario of  80\%/20\%, with gains up to 14\%. Similar to other datasets, the \gls{PBC} is able to correct MixMatch's accuracy impact of data imbalance most of the times, as seen in Table \ref{table:NIS}.

Finally, the test results for  the Indiana dataset are depicted in Table \ref{table:indiana}. The base-line accuracy for the Indiana chest x-ray dataset ranges from 84\% to 88\%. The accuracy gain from implementing MixMatch with \gls{PBC} ranges from 4\% and to 5.6\% versus the base-line supervised model. Implementing the \gls{PBC} versus the original MixMatch implementation yields an  accuracy gain from +4.5\% to +14\%. In the case of this dataset, data imbalance seems to further decrease MixMatch's accuracy, as we can see in Table \ref{table:indiana} when comparing the accuracy results of the 50\%50\% configuration to the 70\%/30\% and 80\%/20\% imbalance settings.

\begin{figure}
\centering{}
\includegraphics[scale=0.28]{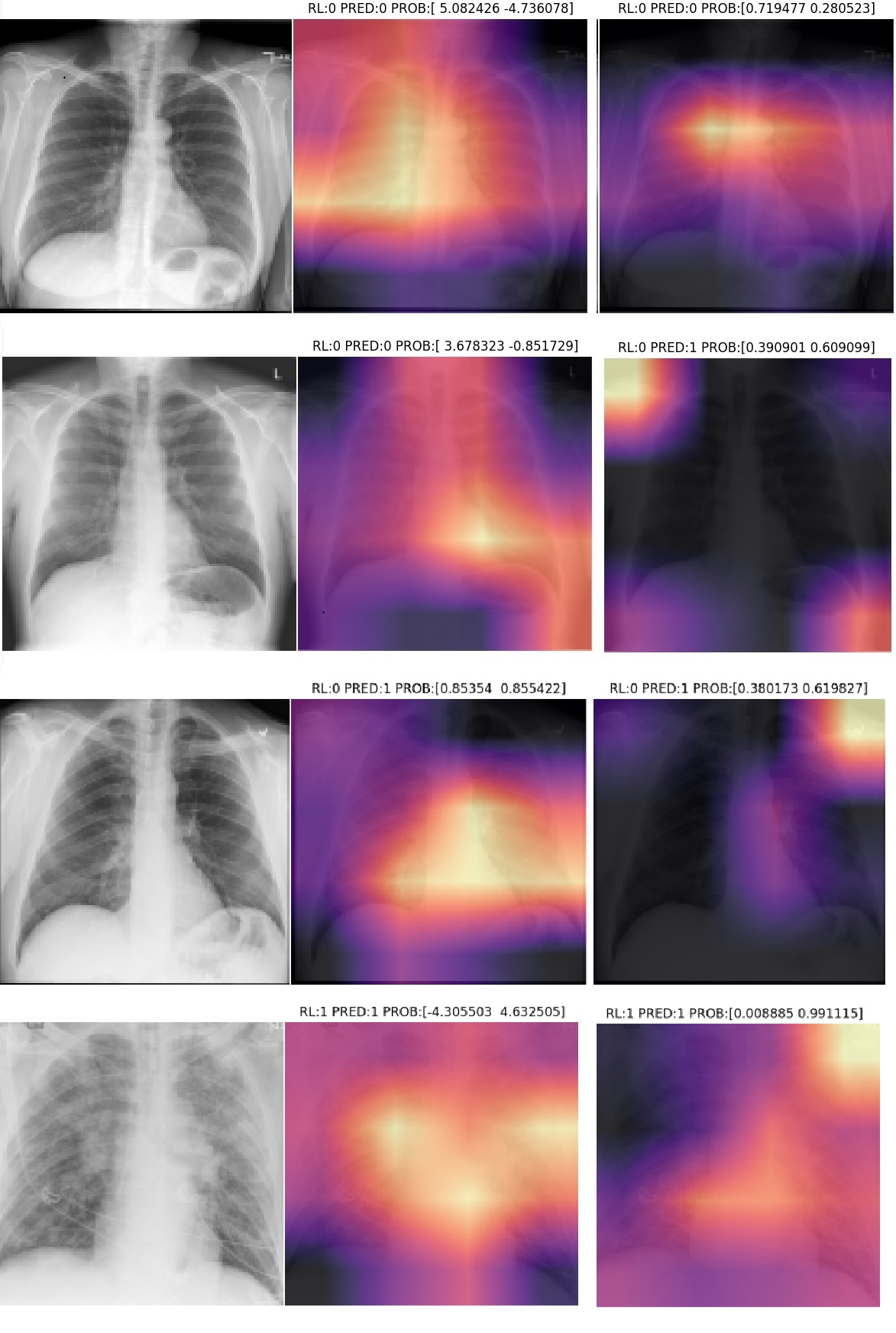}
\caption{ From top to bottom: Two sample heatmaps for the Indiana dataset, and two sample heatmaps for the chest X-ray8 dataset. From left to right:  the original image, the heatmap of the MixMatch trained model with the proposed \gls{PBC} and the output of the supervised model. The legend RL corresponds to the real label, PRED to the model prediction and the array of two values is related to the output net values.
\label{fig:HeatMapsINDIANA}}
\end{figure}

For the tested datasets, the accuracy can be considered to be very similar when evaluating the base-line supervised model under different data imbalance conditions, as seen in Tables \ref{table:costarica}, \ref{table:china}, \ref{table:NIS} and \ref{table:indiana}, suggesting a higher sensitivity of MixMatch when trained with imbalanced data. 
The overall trend of the accuracy gain of using the proposed MixMatch with \gls{PBC} over its original implementation is positive, as seen in \ref{table:summary}, accross all the datasets tested.  Most of the accuracy gains are higher than 3\%, and also most of them are statistically significant, after performing a non parametric Wilcoxon test, with an acceptance criteria of the hypothesis of significant difference between the accuracies of both configurations of $p>0.1$. There are some cases where the default MixMatch implementation does not bring any accuracy gain when facing an imbalanced dataset, as seen for instance in the test results of the Indiana dataset, detailed in Table \ref{table:indiana}. For example the accuracy of the supervised model with 10 labels is around 83\%, and the accuracy of the MixMatch model with no \gls{PBC} is no higher than  83\%. This implies the mandatory need of correcting data imbalancing for the MixMatch model, given its high sensitivity to data imbalance. 
Finally, regarding the qualitative experiments proposed, Figure \ref{fig:HeatMapsINDIANA}  show sample heatmaps for the Indiana and chest X-ray8 datasets, respectively. Both figures reveal how the neural network tends to focus more on lung areas when using the semi-supervised model trained with both datasets. The Densenet121 model trained with MixMatch including the \gls{PBC}  modification yielded an accuracy of 91.3\% for the tested sample from the Indiana dataset, and 67.74\% for the supervised model. For chest X-ray8 dataset, an accuracy of 93.4\% was yielded for the MixMatch framework with \gls{PBC}, and 77.4\% for the supervised model.  We can see in  Figure \ref{fig:HeatMapsINDIANA}   how the hot pixels move towards lung regions when using the semi-supervised model, and also how the net weights of the output layer become steeper. This tends to happen even when the resulting predictions in both models are correct.  

\section{Conclusions}

In this work we have analyzed the impact of data imbalance for the detection of COVID-19 using chest X-ray images. This  is a real-world problem, which can arise frequently in the context of a pandemic, where few observations are available for the new pathology.   To our knowledge, this is the first data imbalance analysis of a \gls{SSDL} designed to perform COVID-19 detection using chest X-ray images.  The experiment results suggest a strong impact of data imbalance in the overall MixMatch accuracy, since results in Table \ref{table:summary} reveal a stronger sensitivity of \gls{SSDL} when compared to a supervised approach. The accuracy hit of training MixMatch with an imbalanced labelled dataset lies in the 2-11\% range, as seen in Tables \ref{table:costarica}, \ref{table:china}, \ref{table:NIS} and \ref{table:indiana}.  This enforces the argument developed in \cite{oliver2018realistic,calderonramirez2020mixmood} which draws the attention upon data distribution mismatch between the labelled and the unlabelled datasets, as a frequent real-world challenge when training a \gls{SSDL} model.  

Moreover, a simple and effective approach for correcting data imbalance by modifying MixMatch's  loss function was proposed and tested in this work. The proposed method gives a smaller weight to the observations belonging to the under-represented class in the labelled dataset. Both the unlabelled and the labelled loss terms were re-weighted, as opposed to the unlabelled re-weighting developed for the mean teacher model in \cite{hyun2020class}, which only modifies the weights of the unlabelled term. This was done since in our empirical tests the unlabelled term had less impact in the overall model accuracy.   For the pseudo-labelled and MixUp augmented observations, we assigned the weights using the pseudo and augmented labels. The proposed method is computationally cheap, and avoids the need of complex and expensive generative approaches to correct data imbalance.  A systematic accuracy gain is yielded when comparing the original MixMatch implementation with the proposed \gls{PBC} for data imbalance correction, as seen in Table \ref{table:summary}. For the tested datsets,  often the proposed  \gls{PBC}  leads to significant accuracy gains from the supervised model, as data imbalance can even hinder any accuracy gain of using MixMatch, as seen in Tables \ref{table:costarica},\ref{table:china}, \ref{table:NIS} and \ref{table:indiana}.  The accuracy gain ranges between  3\% and 11\%, with statistical significance for most of the datasets tested. In most of the datasets, the accuracy gain is higher for the 80\%/20\% imbalance setting. Among the tested datasets, we included a new one with digital X-rays from healthy Costa Rican patients, which we make available for the community. 

This work can be extended by using the customized feature extractors proposed in \cite{cohen2020predicting}, as our architecture uses the more common transfer learning approach from a generic dataset (Imagenet), to later refine the feature extractor. The semantic relevance of the extracted features can be improved along with the model explainability, as seen in Figure \ref{fig:HeatMapsINDIANA}.  However, the proposed solution in this work can be ported to use a more specific feature extractor. Therefore, we plan to test its usage under different customized feature extractors. Furthermore, it is interesting to  investigate the impact of \gls{SSDL} on deep learning explainability/uncertainty  measures. We suspect that unlabelled data can improve models' uncertainty estimations and explainability accuracy.

\section*{Acknowledgments}

This work is partially supported by Spanish grants TIN2016-75097-P, RTI2018-094645-B-I00, UMA18-FEDERJA-084 and the funding from the Universidad de M\'alaga. We acknowledge Clinica Imagenes Medicas Dr. Chavarria Estrada, La Uruca, San Jose, Costa Rica, for its support to the data compilation process of the digital X-ray image dataset used in this work.

\bibliographystyle{plain}
\bibliography{mendeley}

\end{document}